\documentclass[showpacs,prb,preprint]{revtex4}
\usepackage{amssymb}
\usepackage{graphicx}

\begin{document}

\title{Graphene based spin field effect transistor}

\author{Y. G. Semenov and K. W. Kim}
\address{Department of Electrical and Computer Engineering North
Carolina State University, Raleigh, NC 27695-7911}

\author{J. M. Zavada}
\address{U.S. Army Research Office, Research Triangle Park, NC 27709}

\begin{abstract}
A spin field effect transistor (FET) is proposed by utilizing a
graphene nanoribbon as the channel. Similar to the conventional
spin FETs, the device involves ferromagnetic metals as a source
and drain; they, in turn, are connected to the graphene channel.
Due to the negligible spin-orbital coupling in the carbon based
materials, the bias can accomplishes spin manipulation by means of
electrical control of electron exchange interaction with a
ferromagnetic dielectric attached to the nanoribbon between source
and drain. The numerical estimations show the feasibility of
graphene-based spin FET if a bias varies exchange interaction on
the amount around 5 meV. It was shown that the device stability to
the thermal dispersion can provide the armchair nanoribbons of
specific width that keeps the Dirac point in electron dispersion
law.
\end{abstract}

\maketitle

Conventional electronics makes use of the electron charge that
faces major technical limitations in the continued path to device
downscaling. Spintronics\cite{spinel,Wolf} operates by controlling
the electronic spin state. A typical operation of spintronic logic
device or spin-FET transistor as conceived by Datta and
Das~\cite{DattaDas,Gvozdic05} assumes three principle operations:
(1) spin injection into the device active channel, (2) spin
manipulation by the applied bias, and (3) spin detection at the
device drain. In turn, the feasibility of such spin manipulation
imposes some limitations on the semiconductor channel of the
length $L$. The main restrictions are (i) requirement for the long
enough spin-phase relaxation time $T_{2}$ so that spin phase
should not be lost while an electron travels with mean velocity
$v$ through semiconductor channel, i.e. $T_{2}>L/v$ and (ii)
independence of spin phase rotation on electron velocity, i.e. it
should be stable for thermal dispersion of electron energy. Note
that the electric field spin manipulation via Rashba effect
assumes a strong spin-orbital coupling, which conflicts with fast
spin decoherence. Involving the holes with enhance Rashba effect
(compared with electron spin) not only shortens spin coherence but
also breaks a stability to the energy dispersion
because of non-linear spin-splitting dependence on hole velocity.\cite%
{Gerchikov,Winkler,Gvozdic06}

In this work we show that the aforementioned problems can be
avoided in the spin-field transistor based on a different
conception that utilize specific properties of the graphene sheet.
The device design assumes the ferromagnetic source and drain,
which resemble conventional spin transistor (Fig. 1). However spin
rotation occurs in graphene nanoribbon, which provides the
electron exchange interaction with non-metallic ferromagnetic
layer attached to a graphene surface. This exchange interaction
can be treated as an effective magnetic field $G$ so that a
spin-polarized electron
traveling through the graphene during the time $\Delta t$ turns around the $%
G $ on the angle $\theta =G\Delta t/\hbar $. If external electric
field
applied perpendicular to the graphene sheet will change the $G$ by the $%
\Delta G$, the spin rotation varies like $\Delta \theta =\Delta
G\Delta
t/\hbar $ that can perform the spin-valve function at sufficiently large $%
\Delta \theta $.

In order to utilize such spin-manipulation in spin-FET, the system
must meet certain additional criteria. First, a spin coherence
time $T_{2}$ must be long enough compared to traveling time
$\Delta t$, so that decoherence does not influence $\theta $. In
the case of graphene, the extremely small spin-orbital coupling
can fix a long enough $T_{2}\gg \Delta t$ that solves a problem of
spin decoherence. Then, the thermal scattering of the electron
energy at finite temperature $T$ must not influence the spin
rotation. Note that in the case of Datta-Das device, this problem
was solved due to proportionality of spin-rotation rate to
electron velocity that results in compensation of thermal
scattering on electron energy under the quadratic
dispersion law $\varepsilon _{k}\sim k^{2}$. In our case\ the drift time $%
\Delta t=L/v$ inverses to electron velocity $v$ in the channel
under the ferromagnetic layer of the length $L$. So in the case of
the linear
dispersion law in graphene ribbon\cite{Son06,Brey06} the electron velocity $%
v $ does not subjected to thermal scattering that preserves a
broadening of the $\theta $. Finally, the applied bias must
provide a changing of the effective field $G$. Applying the
ferromagnetic materials with giant magnetoelectrical
effect\cite{Fiebig} will provide the spin-rotation sensitivity to
electric field. Consider this possibility in detail.

The problem of spin injection as well as spin detection in
graphene are now in a progress,\cite{Nishioka,Cho} so we focus on
the analysis of spin dynamics in graphene. Although an applied
bias induces some band spin splitting we neglect this effect
hereinafter; besides we neglect the spin-orbital coupling that has
been estimated as extremely small one, $\sim 1 $ $\mu
$eV.\cite{Min06} Graphene consists on honeycomb flat lattice of
carbon atoms with two atoms $A$ and $B$ per unit cell. In the
vicinity of Fermi energy $\varepsilon _{F}=0$ the effective mass
approximation assumes that the spinless wave function takes the
form of a four-vector $(\psi _{A},\psi _{B},\psi _{A}^{\prime
},\psi _{B}^{\prime },)$, where $\psi _{A}$ and $\psi _{B}$ [$\psi
_{A}^{\prime }$ and $\psi _{B}^{\prime }$] are the
amplitudes at the sublattices $A$ and $B$ in the valley $\mathbf{K}=\frac{%
2\pi }{3a}(1,1)$ [$\mathbf{K}^{\prime }=\frac{2\pi }{3a}(-1,1)$]; $a=\sqrt{3}%
a_{C-C}=0.249$ nm is a length of lattice vector, $a_{C-C}$ the
distance between the neighbor carbon atoms. In the vicinities of
valley $\mathbf{K}$
, the Hamiltonian for the envelope wavefunctions takes the form\cite{Ando05}%
\begin{equation}
H_{K}=\hbar v\left(
\begin{array}{cc}
0 & -i\partial /\partial x-\partial /\partial y \\
-i\partial /\partial x+\partial /\partial y & 0%
\end{array}%
\right) ,  \label{eq1}
\end{equation}%
where $v\simeq 10^{8}$ cm/s is a velocity of the electron at Fermi energy.%
\cite{Novoselov} Hamiltonian $H_{K^{\prime }}$ for the
$\mathbf{K}^{\prime }$ valley differs from Eq. (\ref{eq1}) by
opposite in sign $i$. In general case Hamiltonian represents the
direct sum
\begin{equation}
H_{0}=H_{K}\oplus H_{K^{\prime }}.  \label{eq1a}
\end{equation}

Electronic transport will be considered through an armchair
nanoribbon directed along the $y$ axis so that a wave number
$k_{y}$ attributes to a quantum numbers of the electron. The
problem of eigenstates of armchair nanoribbons can be correctly
developed in terms on $kp$ approach with
Hamiltonian (\ref{eq1a}) by applying appropriate boundary conditions.\cite%
{Brey06} One of the remarkable feature of such solutions appears
at particular nanoribbon width $L_{x}=(3p+1)a$ with integer $p>0$
that keeps the Dirac point at zero energy despite the electron
confinement. It can be shown that the energy band structure in
such a case describes\cite{Brey06}
\begin{equation}
E_{m,k_{y}}=\pm \frac{\hbar v}{a}\sqrt{(k_{y}a)^{2}+\frac{(2\pi m)^{2}}{%
(6p+3)^{2}}}  \label{eq2}
\end{equation}%
with integer $m$. The $m=0$ corresponds to gapless linear
dispersion law
alike to the case of bulk graphene; the edge of first excited bands with $%
m=1 $ possesses the energy $E_{1,0}=\frac{2\pi \hbar
v}{3(2p+1)a}$. Note that at reasonably narrow ribbon the energy
$E_{1,0}$ can significantly exceed the thermal energy at room
temperature. For example, at $L_{x}=7.72$ nm ($p=10$) one can find
$E_{1,0}=260$ meV. This estimation shows that the lowest subband
with linear dispersion law can determine an electronic transport
through graphene nanoribbon even at room temperature.

Now let us take into account the electron exchange interaction
with magneto-ordered layer attached to the top side of the
graphene nanoribbon. We assume that a magnetic structure is formed
by the magnetic ions with
non-compensated spin moments $\mathbf{S}_{\lambda ,j}$ localized at sites $%
\lambda $ of the unit cell $j$. Corresponding exchange integrals $J(\mathbf{R%
}_{\lambda ,j})$ crucially depend on overlap of graphene electron
wavefunction with localized magnetic ions; thus the only nearest
neighbor unit cells have to be taken into account when one sums up
over the $j$.
Hamiltonian of exchange interaction takes the form%
\begin{equation}
H_{ex}=\frac{1}{N}\sum_{j=1}^{N}\sum_{\lambda }J(\mathbf{R}_{\lambda ,j})%
\mathbf{S}_{\lambda ,j}\mathbf{s,}  \label{eq3}
\end{equation}%
where $\mathbf{s}$ is an electron spin, $N$ the number of unit
cells attached to nanoribbon. It is naturally to introduce the
molecular fields (in energy units) acted on electron spin in
nanoribbon,
\begin{equation}
\mathbf{G}_{\lambda
}=\frac{1}{N}\sum_{j=1}^{N}J(\mathbf{R}_{\lambda ,j})\left\langle
\mathbf{S}_{\lambda ,j}\right\rangle  \label{eq4}
\end{equation}%
that stem from the exchange interaction with magnetic ions of certain type $%
\lambda $, the $\left\langle \mathbf{S}_{\lambda ,j}\right\rangle
$ is a mean value of spin moment. Futher elaboration of the Eq.
(\ref{eq4}) needs detailed specification of the structure that is
not a scope of this letter.
We only want to stress the dependence of exchange fields $\mathbf{G}%
_{\lambda }$ from both exchange integrals and magnetic moments of
sublattices that will prompt the different mechanisms of $\mathbf{G}%
_{\lambda }$'s manipulation. In the context of this approximation,
the final
expression for the exchange Hamiltonian takes the form of Zeeman energy%
\begin{equation}
H_{ex}=\mathbf{Gs,}  \label{eq5}
\end{equation}%
where $\mathbf{G}=\sum_{\lambda }\mathbf{G}_{\lambda }$ is a total
effective field in energy units. Note that $\mathbf{G}$ can run to
some non-zero amount for any magnetic ordered materials, i.e.
ferromagnetics, ferrimagnetics or even antiferromagnetics if the
magnetic ions, which possess the opposite directed spins are
differently situated with respect to graphene sheet.

The total Hamiltonian is a sum, $H=H_{0}+H_{ex}$, that include the
commutating operators taken from the Eq.(\ref{eq1a}) and Eq.
(\ref{eq5}).
According to Fig. 1, the effective exchange filed is directed along to $x$%
-axis so that the eigenenergy of electron depends on electron spin
projection $\xi =\pm 1/2$ on the axis $x$:%
\begin{equation}
E_{\xi }=\hbar vk_{y}+G_{x}\xi .  \label{eq6}
\end{equation}%
Eq. (\ref{eq6}) shows that an electron of some fixed energy and
with different spin orientations $\xi =+1/2$ and $\xi =-1/2$
possesses a
difference in wave vectors $k_{y}(\xi =-1/2)-k_{y}(\xi =+1/2)=G_{x}/\hbar v$%
. Thus at the end of the graphene nanoribbon of the length $L_{y}$
this produce the spin phase rotation $\theta =G_{x}L_{y}/\hbar v$.
If an applied bias in $z$-direction mediates the change $\Delta
G_{x}$ of the effective field along the whole nanoribbon, the spin
rotation will progressive change,
\begin{equation}
\Delta \theta =\Delta G_{x}L_{y}/\hbar v.  \label{eq7}
\end{equation}%
This result is in accordance of our qualitative discussion. Note that $%
\Delta \theta $ is not depended on $k_{y}$ i.e. it is robust with
respect to electron thermal dispersion.

It can be noted two mechanisms of electric field influence on the
effective field $G$. First, the dielectric polarization in an
external electric field is accompanied by ions shifts inside the
crystal unit cell. If such shift concerns the magnetic ions, it
will result in variation of exchange integrals and, therefore,
exchange field $G$. Other possibility can be realized in the case
of multiferroic materials with giant magneto-electric
effect\cite{Eerenstein,Kumar,Fiebig} where the total magnetic
moment $\sum \left\langle \mathbf{S}_{\lambda ,j}\right\rangle $
is strongly depended on electric polarization.

Now we can estimate the magnitude of the exchange field variation
$\Delta G$ needed to the effective device operation. Note that
spin coherence length in graphene was recently found as long as 1
$\mu $m.\cite{Oezylmaz} Thus, under the applying of graphene
nanoribbon of 100 nm length, the variation $\Delta G\simeq 5$ meV
leads to spin rotation changing $\Delta \theta \simeq 1$. The
spin-controlled current through the device represent the
oscillating function on the $\Delta G$.

Let us introduce the transmission coefficient $T_{+}$ for the
electrons, which reach the drain with spin directed along the
$y$-axis (Fig. 1). Similarly, one can introduce the $T_{-}$ for
anti-directional spin and
magnetization vectors. The total transmission of the device is%
\begin{equation}
T=\frac{1}{2}T_{0}+\left\langle s_{y}\right\rangle \Delta T,
\label{eq8}
\end{equation}%
where $\left\langle s_{y}\right\rangle $ is a mean value of
electron spin projection on the $y$-axis, $T_{0}=T_{+}+T_{-}$,
$\Delta T=T_{+}-T_{-}$. In
the case of ballistic transport $\left\langle s_{y}\right\rangle =\frac{1}{2}%
\cos \theta $. As mentioned above, the spin rotation is $\theta
=\theta _{0}+\Delta \theta $ with $\theta _{0}=(2n+1)\pi $, which
is independent on applied bias and integer $n$. This results in
$\left\langle s_{y}\right\rangle =-\frac{1}{2}\cos \Delta \theta $
so that maximal transmission amplitude $\Delta T$ arrives at
$\Delta \theta $ variation from $0$ to $\pm \pi $. Naturally, the
finite coherence length $L_{C}$ will reduce the $\Delta T$ by the
factor $\exp (-L_{y}/L_{C})$. Fig. 2 illustrates the expected
dependencies of the device transmission on the variable part of
the exchange field $\Delta G$\ and the channel length $L_{y} $.

In summary, a conception of spin-FET based on hybrid structure,
which incorporates the graphene nanoribbon with attached
ferromagnetic dielectric is proposed. The advantage of such hybrid
structure is a long electron spin coherence in graphene and
constant electron velocity through the nanoribbon of specific
width. As a result, the ferromagnetic dielectric can control spin
rotation and implement the function of spin-FET if an effective
exchange field is controlled by applied bias. Apparently the
multiferroic materials are probably most appropriate candidates
for such structure.

This work was supported in part by the MARCO Center on FENA and US
Army Research Office.

\newpage \noindent Figure captions

\vspace{0.2in} \noindent Fig. 1. Schematic illustration of the
spin-FET based on the graphene nanoribbon (circles with bonds) and
ferromagnetic dielectric (FMD) hybrid structure. Ferromagnetic
source\ (S) and drain (D) has collinear magnetic moments (large
arrows) directed with $y$-axis,
ferromagnetic dielectric directs its magnetization (circle with dot) along $%
x $-axis. The figure illustrates the situation when electron spin
(small arrow) inverse its direction due to interaction with
ferromagnetic dielectric that suppresses the conductance of the
device (spin-valve is closed). When an applied bias (along
$z$-axis) $V_{g}$ alters the electron interaction with
ferromagnet, the electron reaches the drain with arbitrary spin
orientation, i.e. it has a finite probability to pass into the
drain with the same spin projection (spin-valve is unclosed).

\vspace{0.2in} \noindent Fig. 2. Device transmission as a function
of (a) exchange field in energy units under the length $L_{y}=100$
nm and (b) a channel length under the $\Delta G=10$ meV. The
coherence length $L_{C}=500$ nm and ratio $T_{-}/T_{+}=0.1$ are
applied for both pictures.

%

\newpage

\begin{center}
\begin{figure}[tbp]
\includegraphics[scale=.6,angle=0]{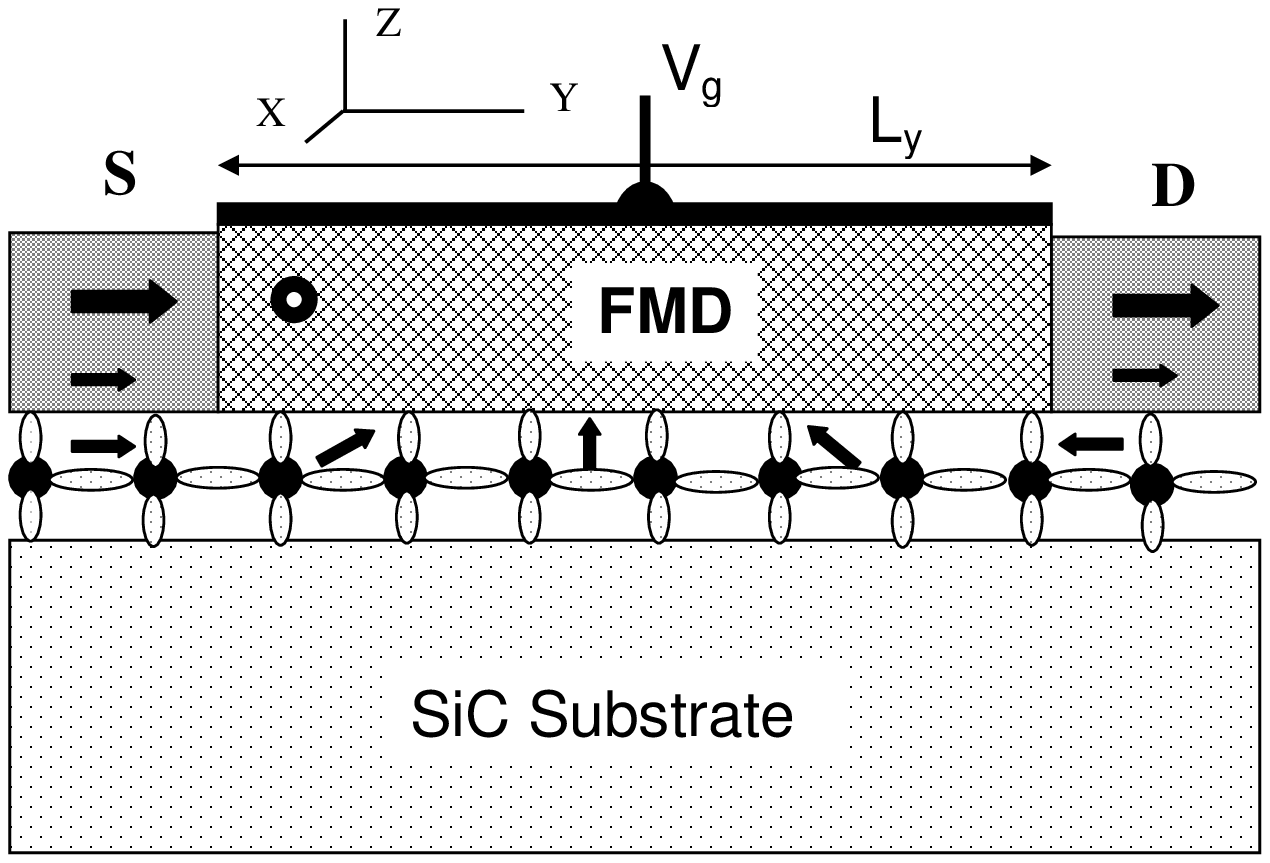}
\end{figure}
\vspace{320pt} {\large Fig. 1: Semenov et al. }
\end{center}

\newpage

\begin{center}
\begin{figure}[tbp]
\includegraphics[scale=.8,angle=0]{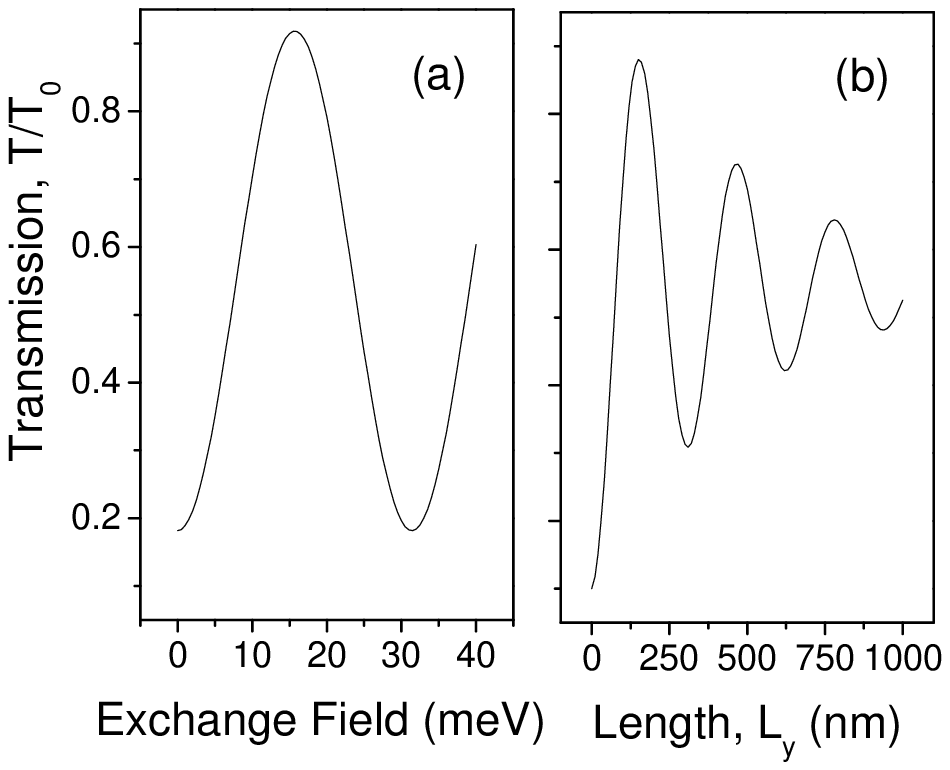}
\end{figure}
\vspace{320pt} {\large Fig. 2: Semenov et al. }
\end{center}

\end{document}